# Laser absorption spectroscopy to probe the superposition principle in atomic translational motion

A.Zh. Muradyan

Yerevan State University, Yerevan, Republic of Armenia

e-mail: muradyan@ysu.am

**Abstract:** It is proposed a few-atom Doppler-sensitive absorption spectroscopy scheme resolving the long-standing dilemma regarding the nature of an atomic quantum translational motion (superpositional or non-superpositional) and its measurement (collapsing or non-collapsing).

## 1. Introduction

The quantum superposition principle is introduced as a direct consequence of the linearity of the Schrödinger equation

$$i\hbar \frac{\partial}{\partial t} |\psi(z,t)\rangle = \hat{H}(z,t) |\psi(z,t)\rangle \tag{1}$$

with respect to the wave function $\psi(z,t)$. It states, from a mathematical point of view, a simple thing that if the wave functions $\psi_1(z,t)$ and $\psi_2(z,t)$ are solutions of equation (1), then any linear combination of them,

$$|\psi(z,t)\rangle = c_1 |\psi_1(z,t)\rangle + c_2 |\psi_2(z,t)\rangle, \tag{2}$$

is also a solution of equation (1). However, its physical assertion is highly concerning, since the wave function is the representative of the state of a physical system at each moment of time. In fact, this principle states that a physical system can be in two (or more) of its possible states simultaneously, either internal or translational. This is in direct contradiction to the classical concept of physics and has never been observed by measurements. For example, if $\psi_1(z,t)$ and $\psi_2(z,t)$ are possible states of a particle with momenta $p_1$ and $p_2$, then it is beyond the classical concept of physics that a particle can have the momenta $p_1$ and $p_2$ together. No particle has ever been experimentally observed to have two or more momenta simultaneously.

This paradoxical situation is circumvented by the introduction of the *postulate* (or *concept*) *of quantum measurement*. It asserts that the result of a measurement can only be one of the eigenvalues of the measured physical quantity. Being an agent of interaction of a physical system with a measuring device, the quantum measurement does not contain dynamic equations of the theory of physical systems, but is described by non-unitary projection operator. It is irreversible and contains probabilistic concept. The postulate specifies that it is precisely the belonging of the measuring device to the macroworld, where cause-and-effect relationships are unambiguous, that determines the departure of the measurement result from the state of superposition to one of its own states. Regarding the postulate assumption, it is noted that the deviation of the measurement result from the superposition state to one of its own states is determined by the belonging of the measuring device to the macroworld, where cause-and-effect relationships are unambiguous.

The problematic nature of the situation can also be attributed to the fact that the methods of conducting measurements register different momentum states of a particle (atom) at different points in space, and therefore it would be natural to realize that the corpuscular part of quantum particle leaves no possibility of registering two or more momentum states simultaneously. The latter would mean simultaneous registering two or more particles. Moreover, the detection of an atom at two or more points contradicts the conservation of energy or mass. So, in order to advance towards clarifying the essence of the concept of quantum measurement of the superpositional state of the translational motion of a particle, one should turn to methods that address different eigenstates in the same spatial region. This could be the method of laser spectroscopy of test radiation, which in [1] was proposed to be used in atom interferometer as the final stage of recording the atomic state, instead of traditional methods of ionization on a hot wire or spontaneous luminescence.

To advance this possibility, Section 2 briefly presents the well-known Kapitza-Dirac (K-D) diffraction [2] in the case of atomic matter wave in the field of near resonant standing wave of laser radiation [3–6], which is regarded in atom optics as a simple and striking example of generation of a superposition state of atomic center of mass motion with an equidistant family of momenta. The question of the applicability of Doppler-sensitive laser absorption spectroscopy as a kind of process for diagnosing the quantum state of the translational motion of an atom generated by a near-resonant standing wave will be theoretically considered in paragraphs 3 and 4. Conclusions will be given in paragraph 5.

The possibility of experimental implementation of the proposed measurement scheme is on the threshold of a modern laboratory for spectroscopy of laser-cooled atomic gases and will allow us to significantly lift the curtain on the problem of quantum measurement and the directly related problem of quantum superposition in the translational motion of atoms.

## 2. Near-resonant Kapitza-Dirac diffraction

K-D diffraction is the scattering of a matter wave by a periodic potential created by a standing light wave, formed by a pair of counter-propagating laser waves. As a result, a family of new matter waves is formed from the incident matter wave, the adjacent momenta of which along the axis of the standing wave differ from each other by the recoil quantum $p_{rec} = 2\hbar k$ of the standing wave, where $\hbar k$ is the momentum of the photon of a traveling wave (fig. 1).

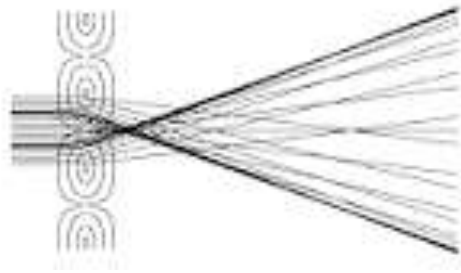

Fig. 1. A simplified diagram of K-D diffraction.

Let us illustrate how this happens by choosing an atom as the object of the matter wave. To ensure the coherent nature of the interaction, the frequency $\omega$ of the standing wave is chosen to be sufficiently far from the frequency $\omega_0$ of the optical transition between atomic energy levels: namely, the resonance detuning $\Delta = \omega - \omega_0$ is much greater than the inhomogeneous width $\gamma$ of the optical transition. Then the motion of the atom occurs with a negligible probability of populating the excited energy level and the atom behaves like a structureless particle.

The solution of equation (1) with a periodic spatially dependent coefficient, according to the Bloch and Fourier theorems, can be written as

$$|\Psi(z,t)\rangle = \sum_{m=-\infty}^{\infty} g_m(t) \exp(i2mkz) |\psi_g\rangle, \qquad (3)$$

where $|\psi_g\rangle$ is eigenstate of the ground energy level, and the probability amplitude $g_m(t)$ is expressed through the Bessel function of the first kind:

$$g_m(t) = J_m(\Omega^2 t / \Delta), \qquad (4)$$

involving the detuning $\Delta$ and the traveling-wave Rabi rate $\Omega$.

Since each exponential in (3) is an eigenfunction of the atomic translational motion with a momentum $p_m = 2m\hbar k$, quantum theory accepts the state (3) as superpositional with respect to these momentum states, requiring that states with different momenta somehow exist before the measurement simultaneously at each moment in time. The reason may be that the Schrödinger equation describes the dynamics of only the wave nature of a quantum object, leaving aside its corpuscular content (a wave packet is not a fundamental concept in the theory). In any case, in the experiments performed so far (see, for example, [7–14]) and in numerous, for example, atom interferometric applications [15–24], only the

$$w_m(t) = |g_m(t)|^2 \tag{5}$$

probabilities of the momentum distribution have been recorded. The results do not contradict, but also cannot serve as justification for the quantum superposition interpretation of (3).

Let us now move on to a discussion of the ability of Doppler absorption spectroscopy of the probe radiation to treat the concept of quantum measurement of states of type (3).

### 3. Probe spectroscopy of the near-resonant Kapitza-Dirac diffracted gas

*Kapitza-Dirac diffract of a 3-level atom.* Since experimental studies and applications of matter wave diffraction most often use alkali metal atoms, we turn to their three-level model with a characteristic hyperfine splitting of the ground level $S_{1/2}$ (see fig. 2).

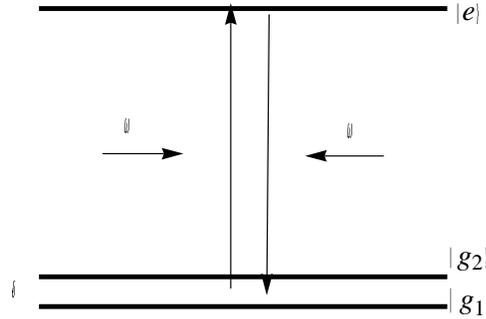

Fig. 2. The scheme of near-resonant K-D diffraction of the 3-level atom matter wave

The Hamiltonian of an atom in the monochromatic standing wave

$$E(\eta,\tau) = E e^{i\eta - i\omega\tau} + E^* e^{-i\eta + i\omega\tau} \tag{6}$$

can be written as

$$\hat{H} = \hat{H}_0 - \hat{d}E(\eta,\tau), \tag{7}$$

where $\eta = kz$ and $\tau = \delta t$ are dimensionless coordinate and time, and $\omega$ stands for $\omega/\delta$.

The general solution (1) for the system under consideration is sought as

$$|\Psi(\eta,\tau)\rangle = g_1(\eta,\tau)|\varphi_{g1}\rangle e^{i\frac{1}{2}\tau-iE_1\tau} + g_2(\eta,\tau)|\varphi_{g2}\rangle e^{-i\frac{1}{2}\tau-iE_2\tau} + e(\eta,\tau)|\varphi_e\rangle e^{-i\Delta\tau-iE\tau}, \qquad (8)$$

where $|\varphi_{g1,2}\rangle$ and $|\varphi_e\rangle$ are the atomic internal state wavefunctions with corresponding energies $E_{1,2}$ and $E$, respectively, and $\Delta$ is detuning of standing wave frequency from the middle of the ground energy levels. Substituting (8) into (1) and applying the standard rotating wave approximation, we obtain the following system of equations for the sought amplitudes:

$$\left(i\frac{\partial}{\partial\tau} \mp \frac{1}{2}\right) g_{1,2}(\eta,\tau) = -2\zeta_{1,2} Cos[\eta] e(\eta,\tau), \qquad (9a)$$

$$\left(i\frac{\partial}{\partial\tau} + \Delta\right) e(\eta,\tau) = -2\zeta_1 Cos[\eta] g_1(\eta,\tau) - 2\zeta_2 Cos[\eta] g_2(\eta,\tau), \qquad (9b)$$

where $\zeta_{1,2} = d_{1,2} E/\hbar\delta$, $d_{1,2}$ is induced dipole moment of optical transition.

Since the coefficients of the system (9a)-(9c) do not depend on time, the general solution is expressed by the exponential functions, namely

$$g_{1,2}(\eta,\tau) = A_{1,2}(1,\eta) e^{-i\lambda(1,\eta)\tau} + A_{1,2}(2,\eta) e^{-i\lambda(2,\eta)\tau} + A_{1,2}(3,\eta) e^{-i\lambda(3,\eta)\tau}, \qquad (10a)$$

$$e(\eta,\tau) = B(1,\eta) e^{-i\lambda(1,\eta)\tau} + B(2,\eta) e^{-i\lambda(2,\eta)\tau} + B(3,\eta) e^{-i\lambda(3,\eta)\tau}, \qquad (10b)$$

where all coefficients $A_{1,2}(...)$, $B(...)$, and $f(...)$ are easily determined analytically and uniquely from the initial conditions of wave function, its first and second derivatives. Due to cumbersomeness, the expressions are not brought in the text.

Now we can move on to the momentum distribution (3) for our research system:

$$g_{1,2}(\eta,\tau) = \sum_{m=-\infty}^{\infty} g_{1,2}(m,\tau)\exp(im\eta), \quad e(\eta,\tau) = \sum_{m=-\infty}^{\infty} e(m,\tau)\exp(im\eta), \qquad (11)$$

determining the coefficients by the inverse Fourier transform:

$$g_{1,2}(m,\tau) = \frac{1}{2\pi}\int_{-\pi}^{\pi} g_{1,2}(\eta,\tau)\exp(-im\eta)d\eta, \quad e(m,\tau) = \frac{1}{2\pi}\int_{-\pi}^{\pi} e(\eta,\tau)\exp(-im\eta)d\eta. \qquad (12)$$

The probability distribution of such momentum states generated for an initially stationary atom on the lower ground energy level is depicted in fig. 3.

The exponents in (11) represent a discrete basis for the translational motion of atoms, implying a quantum superposition entity to the wave function (8).

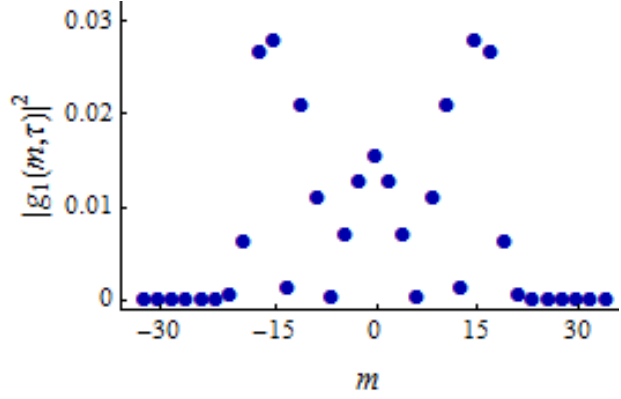

Fig. 3. Probability distribution of momentum states of a 3-level sodium atom due to near-resonance K-D diffraction. Initially, the two hyperfine split ground states are populated equally, and the excited state is unpopulated. $\Delta = 0$, $\varsigma = 10$, and $\omega_r = 10^{-4}$, $\tau = 0.6$.

*Absorption spectroscopy of diffracted atomic gas.* To register and perceive the quantum content arising in an atom as a result of K-D diffraction, we propose to use the method of absorption spectroscopy of probing radiation [25]

$$E_{probe}(\eta, \tau) = E_p e^{ik_p\eta - i\omega_p(\tau - \tau_1)} + E_p^* e^{-ik_p\eta + i\omega_p(\tau - \tau_1)}, \quad \tau \geq \tau_1, \qquad (13)$$

where the amplitude $E_p$ is a slow function of $\eta$ and $\tau$, $\tau_1$ is the moment of switching off the standing pump wave, dimensionless $\omega_p$ and $k_p$ are represented in units of pump running waves. The matter system is an atomic gas so rarefied that the atoms can be considered free before and during interaction with the standing pump wave. To simplify the interpretation of the spectroscopic results to be obtained, it is assumed that the frequency $\omega_p$ of the probe radiation resonates with the dipole-allowed transition from the ground energy levels to some excited level, different from that used in K-D diffraction (fig. 3). Then, for the amplitude $E_p$ of

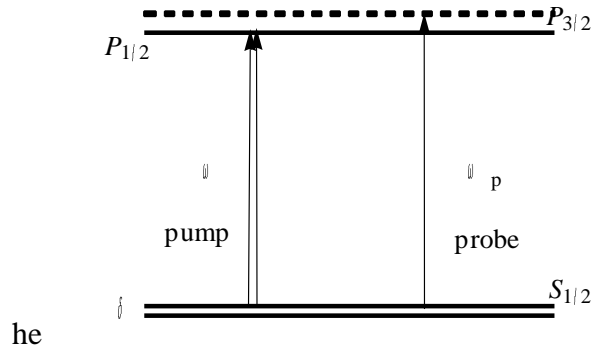

Fig. 4. Interaction scheme of laser absorption spectroscopy. The already absent standing pump wave is presented as additional information that the tested state is just prepared by interaction with this field.

probe wave, the following reduced Maxwell equation [25] is obtained:

$$\frac{\partial}{\partial \eta} E_p(\eta, \tau) = i q' \sum_{m=-\infty}^{\infty} \frac{g_{1,2}(m, \tau_1)}{\Delta_{p,j} - 2m\omega_r k_p - \omega_r k_p^2} g_{1,2}^*(\eta, \tau_1) e^{im\eta} E_p(\eta, \tau) \quad (14)$$

where $q' = 2\pi N |d'|^2 \omega_{1,2}^2 / (\hbar \omega_p \delta)$, $N$ is atomic concentration, $d'$ is induced dipole moment of optical transition, and $g_{1,2}(\eta, \tau_1)$, $g_{1,2}(m, \tau_1)$ are defined by (10a) and (12), respectively.

The solution of (14) at the exit from the medium of thickness $l$ (in $k^{-1}$ units) is

$$E_p(l, \tau) = E_p(0, \tau) \exp[-q' \sum_{m,n=-\infty}^{\infty} \frac{g_{1,2}(m, \tau_1) g_{1,2}^*(n, \tau_1)}{\Delta_{1,2} - 2m\omega_r k_p - \omega_r k_p^2} \frac{e^{i(n-m)l} - 1}{n - m}], \quad (15)$$

where $\Delta_{1,2} = \omega_p - \omega_{1,2}$ is the resonance detuning of the probe field from the corresponding optical transition.

If the thickness $l$ is of the order of or greater than the wavelength of the radiation ($\lambda \sim 10^{-4}$ cm and the number of atoms is huge), then the contribution to the sum over $n$ is practically determined only by the term $n = m$, and accordingly

$$E_p(l, \tau) = E_p(0, \tau) \exp[-i q' \sum_{m=-\infty}^{\infty} \frac{|g_{1,2}(m, \tau_1)|^2}{\Delta_{1,2} - 2m\omega_r k_p - \omega_r k_p^2 + i\gamma} l], \quad (16)$$

where the relaxation parameter $\gamma$ is supplemented phenomenologically.

The subject of our interest, the dependence on the frequency $\omega_p$ enters into (15) and (16) through the detuning $\Delta_{1,2}$. As expected, due to the Doppler effect, the momentum states caused by the standing wave according to (11), (12) and illustrated in fig. 3 are mapped in this spectrum as a series of resonances (fig. 5).

Now we come to the main question of the study: what information can experimental recording of a spectrum (like in fig. 5) give us? The question has two sides. First, whether the atomic state (8) is superpositional in the sense of the simultaneous existence of momentum states, and second, whether the applied measurement collapses the state into one of the basis momentum states. Note, however, that with a huge number of atoms in the sample under study, the spectrum picture will be the same in all four possible cases. For example, if the state is not superpositional, that is, each atom is in one of the basis states (has a certain momentum value), then the probability distribution for the entire sample will again be as in fig. 3 and the measurement of a collapse or non-collapse nature will again form the spectrum of fig. 5. By the way, this is also hinted at by the fact that, unlike formula (15), in formula (16) for a large number of atoms in the sample, the spectrum is determined by the $|g_{1,2}(m,\tau)|^2$ probabilities of the momentum states, and not by the $g_{1,2}(m,\tau)$ and $g_{1,2}(n,\tau)$ amplitudes taken separately.

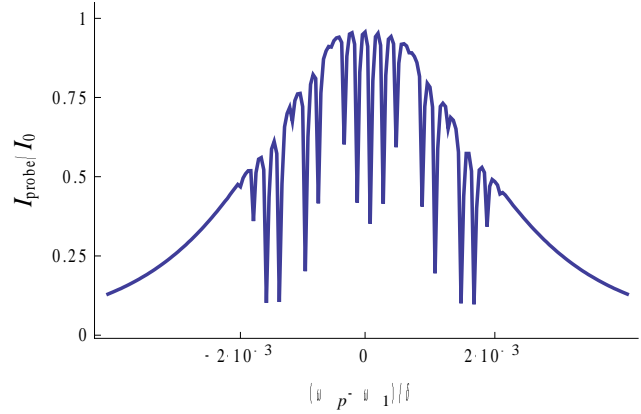

Fig. 5. Absorption spectrum of coherent probe radiation passed through the atomic medium that undergone near-resonance Kapitza-Dirac diffraction. The bell-shaped envelope represents the probe radiation spectrum at the entrance to the medium. $d' = \sqrt{2} \, 10^{-12}$ CGSE, $\delta = 1.77$ GHz, $N = 10^{11}$ cm$^{-3}$, $l = 6\pi$, and $\gamma = 2 \cdot 10^{-5}$ for illustrative purposes.

Therefore, for affirmative answers regarding the nature of the atomic translational motion and the measurement process, it is necessary to make the number of atoms less than the number of momentum states generated by the standing wave, using, for example, the technique of resonator quantum optics or optical lattices. All possible types of experimental absorption spectra in this case are presented in Table 1 below.

Table 1

| State // Measurement | Collapsing | Non-collapsing |
|---|---|---|
| Superpositional | Less number of resonances than in Fig. 5 and equal depths. The spectra under identical initial conditions are different. | The number and relative depths of resonances as in Fig. 5. The spectra under identical initial conditions are the same. |
| Non-superpositional | Less number of resonances than in Fig. 5 and equal depths. The spectra under identical initial conditions are different. | |

As can be seen, the absorption spectrum in three of the possible cases has the same appearance. However, it is precisely in the disputed superposition nature of the translational motion of the atom that the method gives clearly distinguishable spectra for the collapse and non-collapse nature of the measurement.

**4. On the quantum state information formed by a traveling wave following the standing wave**

Information that completes the conclusions from the atomic absorption spectrum can be obtained by exposing the sample to a traveling wave after the standing wave (e.g., from a standing wave composition). Let us first consider the case where the atoms are not exposed to the test radiation field after the standing wave.

The atomic amplitudes are described by equations of the type (9a), (9b), namely

$$\left(i\frac{\partial}{\partial\tau}\mp\frac{1}{2}\right)g_{1,2}(\eta,\tau)=-\zeta e^{-i\eta}e(\eta,\tau), \tag{17a}$$

$$\left(i\frac{\partial}{\partial\tau}+\Delta\right)e(\eta,\tau)=-\zeta e^{i\eta}g_1(\eta,\tau)-\zeta e^{i\eta}g_2(\eta,\tau). \tag{17b}$$

They, as before, are analytically easy to solve, have the same structure as (9a,b)–(12), but with new specific content [26]. The paired superposition of states, one from the ground and one from the excited states of the atom, whose momenta in the traveling wave differ by one photon momentum, leads to a redistribution of the momentum distribution, which at these energy levels oscillates in opposite directions up to macroscopic values. For the lower ground state of the atom under consideration, it is shown in Fig. 6. The corresponding theoretical curve of the atomic absorption spectrum is shown in Fig. 7. It, with a clearly expressed asymmetry, corresponds to the superposition content of the atomic state in interaction with the standing and running pump waves and the "classical" non-collapse nature of the measurement.

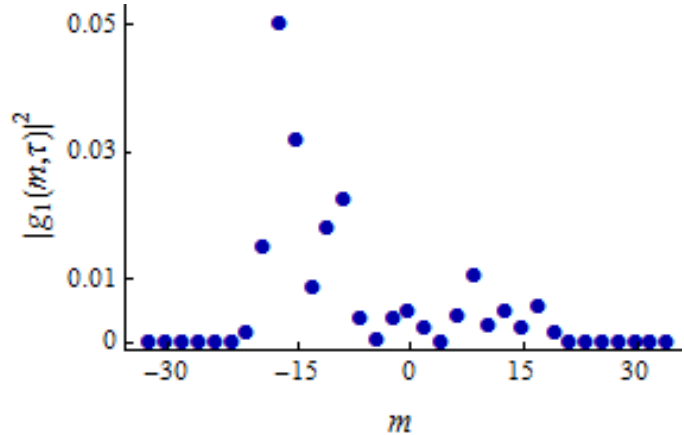

Fig. 6. The atomic momentum distribution formed by a standing and a subsequent traveling wave. The chosen time point $\tau=0.646$ corresponds to the first Rabi flopping of atomic population in the travelling wave.

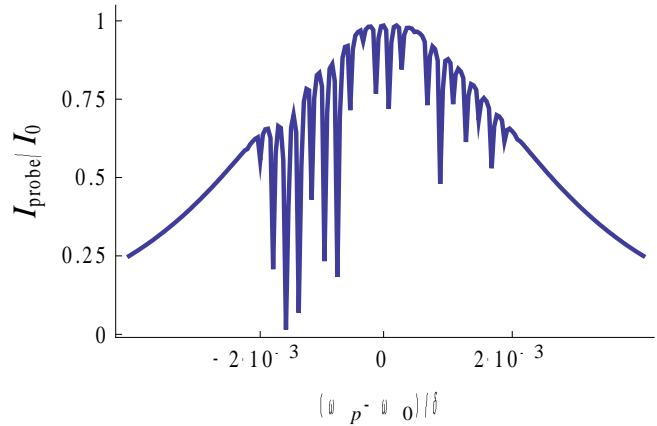

Fig. 7. Absorption of the probe radiation of an atomic sample subjected to the standing and subsequent traveling waves. Atoms were in a superposition state while the measurement was of "classical", non-collapse nature.

All possible outputs of the absorption spectrum with the additional effect of a traveling wave, are presented in Table 2. It is believed that the nature of the atomic state in a standing wave and in the immediately following traveling wave, due to the absence of an intermediate measurement, is the same, superpositional or non-superpositional.

Table 2

| State // Measurement | Collapsing | Non-collapsing |
|---|---|---|
| Superpositional | Less number of resonances, equal and greater depths than in fig. 7. Different spectrum in each experimental realization. | The same number and relative depths of resonances as in fig. 7. Same spectrum in each experimental realization. |
| Non-superpositional | Nearly symmetric distributed resonances, less number, equal depths of resonances for any duration and intensity of the pump traveling wave. | Nearly symmetric distributed resonances, less number, equal depths of resonances. Depth oscillations depending on the duration and intensity of the running pump wave. |

It is noteworthy that the presence of a running pump wave made it possible to separate the collapsing and non-collapsing natures also in the case of a non-superposition nature of the translational state of the atom, which was not possible in the previous consideration of only a standing wave (see Table 1).

Thus, the proposed scheme of Doppler-sensitive measurement of the absorption spectrum of several particles (atoms) clearly separates from each other the possible types of quantum states (superpositional or non-superpositional), as well as the possible nature of the measurements (collapsing or non-collapsing).

If the spectrum of the probe radiation is measured before the effect of the traveling pump wave, then the situation is somewhat similar to that illustrated in Table 1. Namely, the almost symmetrically distributed resonances are of the same depth and smaller in number than in fig. 5. The number of resonances after the standing wave measurement and after the traveling wave measurement is different depending on the duration and intensity of the traveling wave. A new pattern takes place if the states are superpositional and the measurements are non-collapsing. Then, the spectrum after the standing wave repeats all the properties of Fig. 5, while the resonance depths after the traveling wave change periodically, depending on the duration and intensity of the traveling wave.

## 5. Conclusions

Analyzing experiments on recording the translational motion of atoms in near-resonance Kapitsa-Dirac scattering and in an atomic interferometer, it was concluded that they cannot be accepted as a justification for the postulate of the collapse nature of measurement there, as well as the superposition nature of the translational motion of atoms. The blocking factor is that the momentum states are measured at different points in space, where the detection of a superposition state would mean a violation of the law of conservation of energy (or mass), as well as de Broglie's corpuscular-wave concept with a single corpuscle in each particle.

We have proposed and analyzed an experimental method, Dopper-sensitive absorption spectroscopy, which is free from this drawback and, with a sufficiently small number of atoms in the sample, makes it possible to uniquely determine the nature of the translational motion of a particle (superposition or non-superposition) and the nature of the measurement of this state (collapse or non-collapse). Full coverage of the issue is obtained by exposing the atomic sample immediately after the standing wave to a traveling wave and performing the spectroscopic measurement after the traveling one. Possible additional measurement, performed after the standing wave, may be of an auxiliary nature and will increase the experimental reliability of the final conclusions both about the superposition nature of the translational motion of the atom and about the nature of the process of its measurement.